\documentclass[conference]{IEEEtran}
\IEEEoverridecommandlockouts

\usepackage{cite}
\usepackage{amsmath,amssymb,amsfonts}
\usepackage{algorithmic}
\usepackage{graphicx}
\usepackage{textcomp}
\usepackage{xcolor}

\usepackage{mathptmx}
\usepackage{booktabs}
\usepackage{orcidlink}

\usepackage{hyperref}

\usepackage{bfein-defs}
\usepackage{url}
\usepackage{enumitem}

\usepackage{balance}

\def\BibTeX{{\rm B\kern-.05em{\sc i\kern-.025em b}\kern-.08em
    T\kern-.1667em\lower.7ex\hbox{E}\kern-.125emX}}

\makeatletter 
\newcommand{\linebreakand}{%
  \end{@IEEEauthorhalign}
  \hfill\mbox{}\par
  \mbox{}\hfill\begin{@IEEEauthorhalign}
}
\makeatother 

\begin{document}


\title{
  Practical
  Pipeline-Aware
  Regression Test Optimization
  for Continuous Integration
}

\author{
\IEEEauthorblockN{Daniel Schwendner}
\IEEEauthorblockA{\textit{BMW Group}\\
Munich, Germany \\
daniel.schwendner@bmw.de}
\and
\IEEEauthorblockN{Maximilian Jungwirth}
\IEEEauthorblockA{\textit{BMW Group, University of Passau} \\
Munich, Germany \\
maximilian.jungwirth@bmw.de}
\and
\IEEEauthorblockN{Martin Gruber}
\IEEEauthorblockA{\textit{BMW Group}\\
Munich, Germany \\
martin.gr.gruber@bmw.de}
\and
\linebreakand
\IEEEauthorblockN{Martin Knoche}
\IEEEauthorblockA{\textit{BMW Group}\\
Munich, Germany \\
martin.knoche@bmw.de}
\and
\IEEEauthorblockN{Daniel Merget}
\IEEEauthorblockA{\textit{BMW Group}\\
Munich, Germany \\
daniel.merget@bmw.de}
\and
\IEEEauthorblockN{Gordon Fraser}
\IEEEauthorblockA{\textit{University of Passau} \\
Passau, Germany \\
gordon.fraser@uni-passau.de}
}

\maketitle

\begin{abstract}
  %
%
%
Massive, multi-language, monolithic repositories form the backbone of many modern, complex software systems.
To ensure consistent code quality while still allowing fast development cycles, Continuous Integration (CI) is commonly applied.
However, operating CI at such scale not only leads to a single point of failure for many developers, but also requires computational resources that may reach feasibility limits and cause long feedback latencies.
To address these issues, developers commonly split test executions across multiple pipelines, running small and fast tests in pre-submit stages while executing long-running and flaky tests in post-submit pipelines.
Given the long runtimes of many pipelines and the substantial proportion of passing test executions (98\% in our pre-submit pipelines), there not only a need but also potential for further improvements by prioritizing and selecting tests.
However, many previously proposed regression optimization techniques are unfit for an industrial context, because they
(1) rely on complex and difficult-to-obtain features like per-test code coverage that are not feasible in large, multi-language environments,
(2) do not automatically adapt to rapidly changing systems where new tests are continuously added or modified, and
(3) are not designed to distinguish the different objectives of pre- and post-submit pipelines:
While pre-submit testing should prioritize failing tests, post-submit pipelines 
should
prioritize tests that indicate non-flaky changes by transitioning from pass to fail outcomes or vice versa.
To overcome these issues, we developed a lightweight and pipeline-aware regression test optimization approach that employs \rl models trained on language-agnostic features.
We evaluated our approach on a large industry dataset collected over a span of 20 weeks of CI test executions.
When predicting the failure likelihood in pre-submit pipelines, our approach scheduled the first failing test within the first 16\% of tests, outperforming existing approaches.
When predicting test transitions in the post-submit pipeline, it was able to select 87\% of developer-relevant tests by cutting
the test execution time in half and over 99\% within five cycles.

\end{abstract}


\section{Introduction}

Keeping all of an organization's source code in a single repository---i.e., using a \textit{monorepo}---has become a widespread trend in the software engineering industry with Google~\cite{levenberg2016why}, Microsoft~\cite{harry2017largest}, Meta~\cite{goode2014scaling} and Uber~\cite{lucido2017uber} being popular examples.
Even traditional engineering companies such as \bmw are adopting the monorepo design strategy as their products become increasingly defined by software.

While the benefits of monorepos are apparent (\eg avoiding code duplication, clear version pinning, and escaping dependency hell), using a monorepo also implies that a single code change could introduce a breakage that blocks any further development until resolved.
To avoid such breakages that could affect thousands of developers, the usage of continuous integration and regression testing is vital to the successful application of a monorepo.
However, since the number of necessary test executions tends to grow quadratically to the size of the codebase~\cite{Google}, applying CI at such scale creates further challenges in the form of long feedback latencies that decrease developer productivity and high resource consumption that approaches feasibility limits.

\begin{figure}
  \centering
  \scalebox{1}{\input{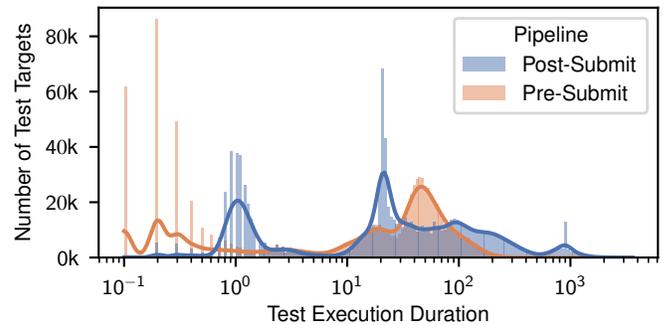}}
  \caption{Comparison of the test target execution duration (seconds) between our pre- and post-submit pipelines.}
  \label{fig:PostTestDurationDistribution}
\end{figure}

To address these challenges, various regression test optimization
techniques have been developed that aim at prioritizing and selecting
test cases that are most likely to
fail~\cite{Ziyuan2011,Srivastava2008}, such that developers can react
more quickly to unblock the CI. However, these techniques are often
not applicable in industrial settings using monorepos---the context
where they are most needed:
First, many techniques rely on complex features such as per-test code
coverage~\cite{Pan2021}, which are not available in practice due to
the significant overhead for measuring them, especially for projects
with high commit frequency~\cite{Google}.
Second, it is common to optimize continuous integration by running
different tests in different pipelines. For example, at \bmw we use
pre- and post-submit pipelines running different tests, where
long-running resource intensive tests are only run in the post-submit
pipeline to not block development (see
\cref{fig:PostTestDurationDistribution}).
Additionally, since permanently broken or flaky tests would block
development in pre-submit pipelines, they are therefore either quickly
resolved or moved to post-submit pipelines.  As a result, we have many
failing test targets (\eg 58\% in our case) in our post-submit
pipelines, which makes selecting tests based on their failure
likelihood infeasible.
Lastly, most techniques are not self-adapting, which would be
important when deployed in a fast evolving environment.
%

We addressed these issues using a lightweight, \rl-based regression
test optimization approach that incorporates pipeline characteristics
into the reward function, but remains generic and language-agnostic by
limiting features to historical execution results and test names.
An evaluation on a large-scale automotive software development project
shows that our approach is effective: In our case study it schedules
the first failing test case within the first 16\% of tests in the
pre-submit pipeline.  In the post-submit pipeline, where developers
are more interested in executing tests that might produce different
results compared to the previous build (i.e., transitions), our
approach selects 87\% of relevant tests within half of the test
execution time for the current code change, and detects 99.78\% of
test transitions within five CI cycles.

The key contributions of this work are:
\begin{description}
  \item[Approach:] We propose a language-agnostic, pipeline-aware, self-adapting regression test optimization approach based on historical test execution results and test names.
  \item[Dataset:] We present an extensive test execution result dataset from a large-scale industry CI system which comprises both pre- and post-submit pipelines.
  \item[Evaluation:] We evaluate our approach on the dataset, showing that it outperforms existing techniques.
\end{description}

All data is provided openly\footurl{https://github.com/code-byter/pipeline-aware-regression-test-optimization}.


\section{CI Infrastructure and Challenges at BMW}
\label{sec:ci_infra}

Our monorepo contains code responsible for driving dynamics and autonomous driving software.
It consists of approximately 70 million lines of code spanning multiple programming languages, most prominently C++ (production code) and Python (tooling code).

\subsection{Our CI Infrastructure}

\begin{figure}
    \centering
    \includegraphics[trim={1mm 0 0 0},clip]{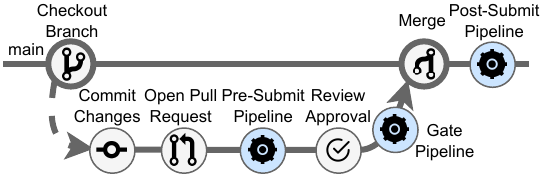}
    \caption{Overview of the executed CI pipelines in the software development process. For each code change, Zuul executes three pipelines. The check and gate pipelines are executed pre-submit, whereas the post pipeline is executed post-submit.}
    \label{fig:pipeline_overview}
\end{figure}

We use Zuul~\cite{Zuul} as our CI system and Bazel~\cite{Bazel} as the build and test tool.
Zuul is based on the concept of pipelines.
A pipeline represents a workflow consisting of jobs planned for execution and is triggered by specific events, with each job performing a specific task such as executing a test suite.
A test suite contains multiple Bazel test targets, consisting of one or more test cases.
The type of test case can vary from unit tests to system-level tests.
Different pipelines are executed at different times, as shown in~\cref{fig:pipeline_overview}, with the following characteristics:

\begin{itemize}
    \item \textbf{Pre-submit pipelines} run before a change is merged into the main branch.
      These pipelines run fast, deterministic tests to provide quick feedback to developers, \eg unit or small component tests.
    \item The \textbf{Gate pipeline} runs after the pull request (PR) is approved.
      It ensures compatibility between concurrent changes by performing a speculative merge of multiple PRs and testing the resulting software again.
      After a PR has passed this pipeline, it is automatically merged into the main branch. This pipeline runs the same tests as our pre-submit pipelines. We did not apply our regression test optimization, as we want to have a fail-safe mechanism (\ie still revealing a regression spotted before merging).
    \item \textbf{Post-submit pipelines} are triggered after a PR is merged into the main branch.
      These pipelines execute tests that take too long or behave too flaky to be executed pre-submit, \eg hardware in the loop tests or simulations testing our autonomous driving capabilities.
\end{itemize}

On a daily basis, our \num{6500} developers trigger approximately \num{10000} pipeline runs, causing 28~million test executions with a peak resource consumption of \num{37000} vCPUs running concurrently.
The size and diversity of our codebase produce substantial challenges, making it infeasible to build everything from scratch ($\sim$6h runtime on a 32-core machine) and run all tests against every change.
Therefore, we identify changed files in each PR and use Bazel to identify affected build and test targets.
We identify related targets, such as targets that transitively depend on modified files in the build graph, have modified files as inputs or contain downstream dependencies to directly affected targets.
This approach heavily reduces load on our CI, however, our pre- and post-submit pipelines still have substantial execution times of 34min resp.\ 56min on average from triggered to finished.
Long pipeline durations are directly related to higher development costs, both through computational resource consumption, and delayed feedback for developers, motivating the need for further optimizations.

A prominent way to increase CI efficiency is \textit{regression test optimization}, which aims at executing relevant tests first (test prioritization) and skipping irrelevant ones (test selection).
Applying this to a large-scale monorepo and CI infrastructure such as ours, however, comes with several challenges.

\subsection{Optimization Challenge: Size and Diversity}

Given the sheer size and diversity of our codebase, traditional code coverage features and complexity metrics are unavailable at the test target level, making it infeasible to leverage them for regression test optimization.
Additionally, regression test optimizations that use language-specific features would require substantial engineering effort and continuous adaptation as new languages are introduced or existing ones evolve.
Consequently, any regression test optimization technique aimed at our CI systems must be \textit{language-agnostic} and \textit{lightweight}.

\subsection{Optimization Challenge: Rapidly Evolving Environment}

On a daily basis, our developers merge over \exnum{500} pull requests.
Apart from source code changes, these pull requests also change the test code, constantly adding new tests, or rewriting or removing existing ones.
A regression test optimization technique operating in this environment should be \textit{self-adapting} to these constant changes.

\subsection{Optimization Challenge: Pipeline-awareness}

Regression test optimization aims at executing relevant tests first (test prioritization) and skipping irrelevant ones (test selection).
The definition of what a relevant tests is, however, depends on the CI pipeline in which the tests are executed:

In pre-submit pipelines, a test is relevant if it fails.
Test selection techniques should therefore aim to skip passing tests.
Test case prioritization can also be applied to optimize pre-submit pipelines:
The goal of a pre-submit pipeline is to decide whether a certain code change can be integrated into the main branch or if it should be blocked as it contains a regression.
This question can be answered after a first test failure has been found without executing all test cases, allowing for a fail-fast test execution strategy, which would benefit from failing tests being executed first.

For post-submit pipelines, test executions are relevant if they deliver an outcome different from the CI cycle of the previous code change (i.e., test transitions).
This means that---contrary to what most evaluations of regression test optimizations assume---a test failure in a post-submit pipeline is not relevant per se, since the test might have already been broken before the code change was merged.
Only if the test fails or passes \enquote{freshly} (i.e., transitions), its result is relevant.
Furthermore, test prioritization is far less relevant in post-submit pipelines compared to pre-submit pipelines:
If a PR causes a post-submit test to transition from passing to failing, the PR needs to be reverted or the regression needs to be fixed.
Since both require manual (review) effort, a $\pm$60min feedback latency (which is approximately the runtime of our post-submit pipelines), does not make a notable difference.
Furthermore, since our post-submit pipeline also contains flaky tests, reverts are anyways not triggered after a single test failure, but only after several consecutive failures.

In conclusion, regression test optimization techniques have to consider the different goals and characteristics of pre- and post-submit pipelines, \ie they have to be \textit{pipeline-aware}.

\subsection{Optimization Challenge: Test Flakiness}

Non-deterministic test behavior (test \textit{flakiness}~\cite{luo2014empirical,parry2021survey}), is a major impediment to continuous integration, as it undermines the central assumption of regression testing, which is that every test failure is caused by the most recent code changes.
Flakiness is especially problematic in pre-submit pipelines, because it blocks developers from merging their pull requests, forcing them to waste their time investigating test failure that are unrelated to their changes~\cite{gruber2022survey,leinen2024cost}.

\begin{figure}
  \centering
  \includegraphics[trim={2.3mm 0 0mm 0},clip]{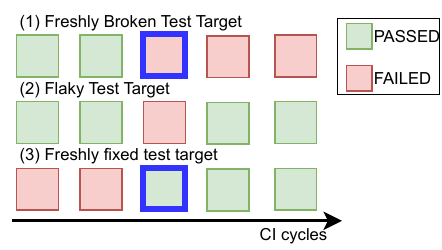}
  \caption{Test execution results over consecutive CI cycles (post-submit pipelines). Only the blue marked test results are relevant for developers, as they indicate changes in code quality.}
  \label{fig:post-submit}
\end{figure}

To mitigate the impact of test flakiness, our team has developed multiple mitigation strategies that
(1) automatically rerun failing tests if the failure is suspected to be caused by an infrastructure issue, or if the test has shown flaky behavior in the past (operating in both pre- and post-submit pipelines),
(2) detect flaky tests in pre-submit pipelines by rerunning all tests on idling, on-premise resources at night to spot flaky test targets, similar to other industry practitioners~\cite{kowalczyk2020modeling,hoang2024presubmit}, and
(3) automatically propose moving flaky pre-submit tests to the post-submit pipeline, where regressions can be more easily distinguished from flaky failures, as they cause consecutive test failures~(\cref{fig:post-submit}).
This process helps us to maintain low flakiness rates in our pre-submit pipelines, however, it increases flakiness in post-submit stages.

Previous work has shown that flakiness can have devastating impact on regression test optimization techniques~\cite{Google,Elbaum2014,fallahzadeh2022impact}.
Therefore, any regression test optimization technique aiming to be applicable in our CI systems has to take test flakiness in post-submit pipelines into account by considering non-flaky test transitions from passed to failed (regression) or vice-versa (bug fix) as relevant, \ie to be executed.




\section{Approach}

\begin{figure}[t]
  \centering
      \includegraphics[trim={0 0 2.5mm 0},clip]{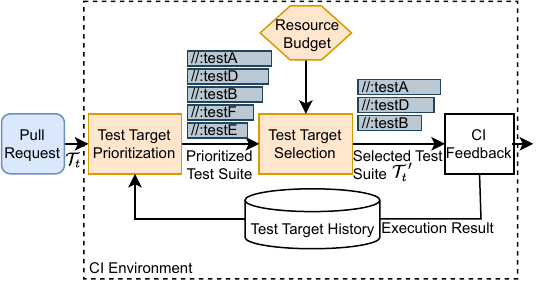}
  \caption{Test target prioritization and selection process.}
  \label{fig:CheckRLEnv}
\end{figure}

Our regression test optimization approach (\cref{fig:CheckRLEnv}) employs \rl for ranking test targets based on pipeline-dependent reward functions and selecting the highest ranked tests until a given resource budget (\ie execution~time) is reached.
Basing the selection process on the results of the prioritization allows it to optimize against the same goal.

\subsection{\rl for Regression Test Optimization}
Its capabilities for continuous, feedback-based learning make \rl (RL) an ideal choice for implementing regression test optimization techniques operating in fast changing CI environments.
To build an RL model, one has to define an \rlagent that interacts with the \rlenv.
The agent learns how to take actions that lead to desirable outcomes by interacting with the environment.
It selects actions (test prioritizations) based on observations of the environment and receives rewards as feedback to guide its learning.

\cref{fig:CheckRLAlgo} shows an overview of the approach.
We employ \textit{pointwise} ranking~\cite{liu2009learning}, meaning that our \rlagent assigns an individual score to each test target independently ($\{a_1, \ldots, a_N\}$), which resembles the relevance of its execution.
We use these scores to rank our test targets (test prioritization).
The environment (\ie~the CI system) then executes the highest ranked tests according to the agent's prioritization until reaching a defined resource budget (test selection).
The observations reported back to the agent ($\{s_1, \ldots, s_N\}$) contain preprocessed information about each test, including historical execution metrics and results (\cref{table:inputfeatures}).
Rewards ($\{r_1, \ldots, r_{N'}\}$) are allocated to the \rlagent for each scheduled and executed test target independently.
This feedback, consisting of initial state, action taken, assigned reward, and resulting state ($<s_i, a_i, r_i, s_{i+1}>$) is then fed back to the agent for further training.
To implement the agent, we use \deepq~\cite{Watkins1992, Mnih2015, Tan2017}, a popular RL method.



\subsection{Input Features}
\begin{table}
  \caption{Description of input features.}
  \centering
  \begin{tabular}{ lp{52mm} }
      \toprule
   \textbf{Feature Name} & \textbf{Description} \\
   \midrule
   {\verb|test_result[]|} & Ordinal encoded test results of the last $k$ executions\\
   {\verb|test_name_PCA[]|} & Test target name preprocessed using a Bag-of-Words (BoW) model with Principal Component Analysis (PCA)\\
   {\verb|last_failure|} & Number of CI cycles since the last failure of the test target\\
   {\verb|last_execution|} & Number of CI cycles since the last execution of the test target\\
   {\verb|avg_duration|} &  Previous average execution duration of the test target\\
   \bottomrule
  \end{tabular}
  \label{table:inputfeatures}
\end{table}

\cref{table:inputfeatures} shows the set of input features that our approach uses for every individual test target.
To ensure applicability even in large-scale monorepos, we only employ language-agnostic and easily retrievable features.

{\verb|test_result[]|} encodes the last $k$ test verdicts in ordinal form of the given test target, where a passing test is represented as 0, and a failure as 1. We determined $k=25$ empirically.

{\verb|test_name_PCA[]|} encodes the test target name using a Bag-of-Words (BoW) model combined with Principal Component Analysis (PCA) for dimensionality reduction.
A test target name is a fully qualified path from the root of the monorepo to the test target.
Due to our naming conventions, this name encodes relevant information, such as affected software components and the test type, \eg~{\verb|//app/componentA/feature1:unit_tests|} \mbox{representing} our naming scheme in anonymized form.
The BoW model represents the test target name as a vector of word counts.
However, high-dimensional sparse vectors are not suitable for machine learning models.
Therefore, we apply PCA to reduce the dimensionality of the BoW vectors, capturing the most important information while reducing noise and redundancy.
Processing source file names using a BoW model combined with PCA is a viable approach to include source file-related information~\cite{Abdelkarim2022}.
BoW models are widely used in natural language processing for representing texts in a numerical format, as they enable capturing topic and correlation information~\cite{Sethy2008}.
PCA is a suitable option for dimensionality reduction, as it produces better results than other methodologies like Latent Semantic Analysis~\cite{Deerwester1990,Ljungberg2017}.

{\verb|last_failure|} represents the number of CI cycles since the last failure of the test target. This feature provides information about the recent stability of the test target.

{\verb|last_execution|} represents the number of CI cycles since the last execution of the test target. This feature indicates how recently the test target was executed, which is relevant to prioritize test targets that have not been executed recently.

{\verb|avg_duration|} represents the previous average execution duration of the test target. This feature is useful for prioritizing shorter tests when resources are limited, or for balancing the scheduling of longer tests when beneficial.

By using this concise feature set, we aim to minimize preprocessing overhead while still capturing relevant information for test target prioritization. The combination of test results, test target names, failure history, execution recency, and execution duration provides a comprehensive yet lightweight set of features for our approach.

\subsection{Pipeline-Dependent Reward Functions}

\paragraph{Per-submit pipelines}
Commonly used reward functions for regression test optimization do not incorporate costs and developer feedback time~\cite{Lima2022, Spieker2017}. To overcome those drawbacks we propose the \costrank function for ranking pre-submit test targets. It incorporates the rank and execution costs as the time elapsed between the start of the execution of the first test $t_1$ and the current one $t_i$. The cost $c_n$ of the test target $t_n$ is the execution duration and $c_{T_t} = \sum_{n=1}^{|\mathcal{T}_t|}{c_n}$ denotes the execution cost of the scheduled full test suite $\mathcal{T}_t$. Further, $\mathrm{rank}(t_i)$ returns the position of $t_i$ in the prioritized suite $\mathcal{T}_t$.
\begin{equation}
    \textit{CostRank}(t_i) =
    \begin{cases}
        \,\,\,\,1 - \alpha \frac{\sum_{n=1}^{\mathrm{rank}(t_i)-1}{c_n}}{c_{T_t}}, &  \text{if } \mathrm{fails}(t_i)\\
        -1 + \alpha \frac{\sum_{n=1}^{\mathrm{rank}(t_i)-1}{c_n}}{c_{T_t}}, & \text{else}
    \end{cases}
\end{equation}
The factor $\alpha$ weights the importance of the cost factor in the reward calculation. We empirically determined $\alpha = 0.9$ to work best in our use case. The reward is positive for failing test cases and negative for passing ones. The value of the reward ranges from -1 to 1, with higher rewards for failing test cases executed early and lower rewards for failing test cases executed later. If a failing test case is executed at first, the reward is $1$, whereas a failing test case ranked last in $T_t'$ receives a reward of $1 - \alpha$. An early scheduled test case reporting a failure with a low execution time returns the highest reward value. This approach reinforces scheduling failing tests earliest to reduce feedback delays for developers, which is backed by our empirical evaluation of pre-submit dataset.

\paragraph{Post-submit pipelines}
Based on the importance of developer-relevant transitions (see \cref{sec:ci_infra}) in post-submit pipelines, we propose the \costcrank function to rank test targets, based on execution costs and developer-relevant transitions.
\costcrank incorporates test cost by returning the negative normalized execution duration $c_i$ for unchanged test results and non-developer-relevant transitions, which discourages the execution of (longer) irrelevant test targets.

\begin{equation}
    \textit{CostChangeRank}(t_i) =
    \begin{cases}
        -1, & \text{if } \mathrm{flaky\_transition}(t_i) \\
          \,\,\,\,\,\,\,1, & \text{else if } \mathrm{relevant\_transition}(t_i)\\
        - c_i,             & \text{otherwise}
    \end{cases}
\end{equation}
We reward the selection of developer-relevant transitions (defined in \cref{sec:ci_infra}) by returning $1$ and especially discourage the selection of flaky transitions by assigning $-1$ to accommodate for the devastating impact that test flakiness can have on regression test optimization techniques~\cite{Google,Elbaum2014,fallahzadeh2022impact}.

\subsection{Model Characteristics and Training}
\begin{figure}
  \centering
      \includegraphics[trim={3mm 0 3.5mm 0},clip,width=\linewidth]{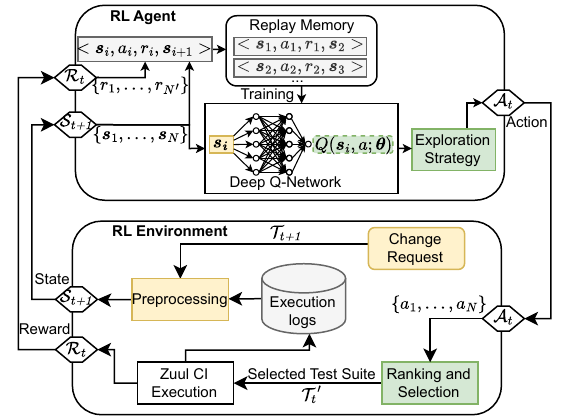}
  \caption{Overview of our regression test prioritization and selection infrastructure: The RL agent estimates a priority score being the state-action values using a deep Q-network. The RL environment schedules the test suite $\mathcal{T}_t'$ and provides a reward based on the test results. Historical CI execution results are stored in a database.}
  \label{fig:CheckRLAlgo}
\end{figure}
The backbone of our \rlagent is \deepq (see \cref{fig:CheckRLAlgo}) using a model consisting of four fully-connected layers with dropout layers with L2-regularization to mitigate overfitting to an old state of the \rlenv.
We initialized our model randomly and trained it using ADAM~\cite{Kingma2014} based on the earlier defined reward functions.

To maintain sample efficiency and stability during training we employ experience replay~\cite{Fedus2020, Lin1992, Mnih2015}. This mechanism involves storing experiences in a fixed length replay buffer (\exnum{4096} in our case). If this buffer is filled, we uniformly sample batches of experiences for training. We first initialize the buffer with historic examples to overcome the problem of starting without any knowledge.

To adjust to the changing environment, the agent must keep a balance between exploitation and exploration.
A mitigation is to use the $\epsilon$-greedy exploration strategy~\cite{Mnih2015}.
However, as our approach relies on continuous action values, this cannot be applied.
Therefore, we employ a similar strategy as RETECS~\cite{Spieker2017}, by drawing a random value from a Gaussian distribution and adding it to the Q-Values.



\section{Case Study}
The proposed approach builds upon the earlier shown processes and optimizations of our CI system (see \cref{sec:ci_infra}) to further reduce developer latencies while decreasing load on our CI.
To assess the effectiveness of our approach, we conducted a case study aiming to answer the following research questions using CI replays to not harm the production environment:
\begin{description}
  \item[RQ1 (Pre-Submit):] How well can our approach reduce feedback latency
    and resource consumption
    of pre-submit testing pipelines?
    (test prioritization and selection)
  \item[RQ2 (Post-Submit):] How well can our approach reduce resource consumption
    of post-submit testing pipelines while selecting transitioning test cases?
    (test selection)
\end{description}

\subsection{Baselines}
For RQ1 we tested our reward function against the commonly used \rnfail function~\cite{Spieker2017,PradoLima2020}, to see how our tailored reward function can improve against our set objectives.
\begin{equation}
  \mathrm{RNFail}(t_i) =
  \begin{cases}
      1, & \text{if } \mathrm{fails}(t_i)\\
      0,             & \text{otherwise}
  \end{cases}
\end{equation}
It returns 1 if a test target fails, otherwise 0.
Furthermore, we chose three baselines to compare against our approach for pre-submit pipelines. The first baseline algorithm is RANDOM, it samples uniformly from all test targets. We repeated that baseline \exnum{10000} times and took the average.
This indicates the performance of a test case selection without any prior ranking, such as the current setup.
The second baseline approach is ROCKET \cite{Marijan2013}, as it has been shown to outperform the RANDOM approach.
ROCKET prioritizes test cases without considering the execution time and selects them based on their rank and available budget.
Further it provides a comparison between machine learning based models and heuristics that utilize historical test information.
Our third baseline is COLEMAN, a \rl algorithm, which is state-of-the-art \cite{Lima2022}. We use the publicly available open-source implementations of COLEMAN and implemented ROCKET ourself due to the missing replication package.

To assess how our approach can be applied to post-submit regression testing (RQ2), we first compare it against a random-baseline. RANDOM uniformly samples from all test targets (repeated \exnum{10000} times and took the average). Furthermore, we compare against the simpler \rnchange reward function.
\begin{equation}
      \mathrm{RNChange}(t_i) =
      \begin{cases}
          -1, & \text{if } \mathrm{flaky\_transition}(t_i) \\
          \,\,\,\,\,\,1, & \text{else if } \mathrm{transition}(t_i)\\
          \,\,\,\,\,\,0,             & \text{otherwise}
      \end{cases}
  \end{equation}
It assigns a reward of 1 to developer-relevant transitions, -1 to flaky transitions, and 0 otherwise.

\subsection{Datasets}

We collected two datasets from our pre-submit and post-submit pipelines' execution logs over a span of 20 weeks.
The dataset contains only non-aborted cycles resulting in at least one failed test target, as otherwise biases are introduced~\cite{Bagherzadeh2022}. For example, choosing 1 as value for NAPFD (\cref{sec:metrics}) for cycles without any failures during evaluation, indicating ideal ranking, leads to misleading results~\cite{Bagherzadeh2022}, especially when reporting the average values across all cycles. Furthermore, we ignored all cycles containing less than six test targets. For such small test suites, the NAPFD metric leads to high values, thus boosting the results~\cite{Bagherzadeh2022}.

Our test execution reports can have one of the following statuses: PASSED, FAILED, FAILED TO BUILD, NO STATUS, TIMEOUT, or FLAKY. We ignore the NO STATUS, TIMEOUT and FAILED TO BUILD results because they do not provide any indication of the test outcome. We consider a test target with FLAKY as passed, as the test passed within its set of intermediate reruns. The post-submit dataset does not contain any FAILED TO BUILD as our CI enforces that our main branch can be built.

\begin{table}
  \caption{Pipeline characteristics and test target results.}
  \begin{center}
      \begin{tabular}{ lrr }
       \toprule
        & \textbf{Pre-Submit} & \textbf{Post-Submit}  \\
       \midrule
       \textbf{Characteristic} & & \\
       \midrule
          Failed Targets (\%) & \num{2} & \num{58.3} \\
          Unique Targets & \exnum{10222} & \exnum{363}  \\
          Avg. Target Duration (sec.) & \exnum{6.53} & \exnum{75.14} \\
        \midrule
        \textbf{Test Results} & & \\
        \midrule
        PASSED & \exnum{947407} & \exnum{413866}\\
        FLAKY & \exnum{298} & \exnum{2184} \\
        FAILED & \exnum{19247} & \exnum{613553}\\
        TIMEOUT & \exnum{30} & \exnum{9426} \\
        FAILED TO BUILD & \exnum{427} & - \\
        NO STATUS & - & \exnum{13708} \\
       \bottomrule
      \end{tabular}
      \label{table:postCharacteristicsComparison}
  \end{center}
\end{table}

The pre-submit dataset consists of 1766 failed cycles, with a CI cycle containing a median of 450 non-cached test targets.
The majority of test target executions were successful ($> 97\,\%$), while less than $2\,\%$ failed, see \cref{table:postCharacteristicsComparison}.

Our post-submit pipeline is executed after merging, resulting in fewer executions compared to pre-submit but with more failed runs within the 20 weeks (\exnum{10752}).
Since the post-submit pipeline is executed after a change has been merged, developers are not blocked by failing post-submit test targets, which lead to permanently broken tests (\num{58.3}\% of all post-submit test executions fail).
The post-submit pipeline executes less targets compared to pre-submit (\num{363} vs. \num{10200}), but with higher average execution times (\num{75.14}s vs. \num{6.53}s).
The distribution of execution times of the test targets within the different pipelines can be seen in \cref{fig:PostTestDurationDistribution}.
%
%
In our post-pipeline dataset, \num{2.96}\% of non-cached test executions result in a change of the test outcome (\ie a test transition).
Test result changes are defined as all executions with a different result compared to the last execution of the respective target, including transitions based on flakiness or other external factors.


We made our datasets publicly available\footurl{https://github.com/code-byter/pipeline-aware-regression-test-optimization} (the test target names have been anonymized).

\subsection{Test Flakiness}

\begin{figure}
  \centering
  \scalebox{1}{\input{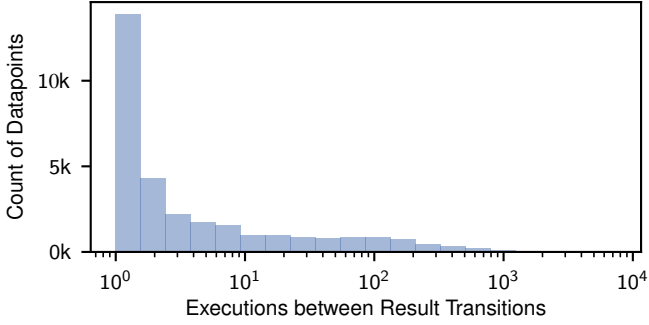}}
  \caption{Distribution of the number of test executions between two test result transitions.}
  \label{fig:PostChangeFrequency}
\end{figure}

To apply the $\textit{CostChangeRank}$ reward function, one has to distinguish flaky from non-flaky test transitions.
Following other industry practitioners~\cite{Leong}, we regard rapidly changing transitions as flaky.
Namely, we define a flaky transition as one that switches signals again within three consecutive CI cycles.
We chose this threshold based on our churn rate (\ie the number of changes merged per time interval) and the time it takes a developer to resolve a regression that has already been merged on the main branch.
Most of our tests transitions switch their signal again within the chosen threshold as can be seen in \cref{fig:PostChangeFrequency}, indicating flaky transitions.

\subsection{Evaluation Metrics}

\label{sec:metrics}

Test case selection aims to execute only relevant test cases.
For pre-submit pipelines, relevance means the effectiveness in detecting code faults.
We assume every test case reveals only one fault of the same severity.
We use the normalized average percentage of faults detected (NAPFD) metric to evaluate the failure detection performance, as it considers both the effectiveness of finding failures in a selected test suite $\mathcal{T}'$, and the rank of the detected test failures~\cite{Qu2007}.
Further, it is well-suited for test case selection, as it accounts for undetected faults~\cite{Lima2022}.
The computation of the NAPFD requires knowledge of the execution results for the full test suite $\mathcal{T}_t$, which is available in all used datasets.

\begin{equation}
    \mathrm{NAPFD}(\mathcal{T}_t') = p - \frac{\sum_{t \in {\mathcal{T}'}^{\text{fail}}_t}{\mathrm{rank}(t)}}{\mid {\mathcal{T}'}^{\text{fail}}_t \mid \cdot \mid {\mathcal{T}'}_t \mid} + \frac{p}{2\cdot \mid {\mathcal{T}'}_t \mid}
\end{equation}

$ {\mathcal{T}'}^{\text{fail}}_t $ is the set of failed tests of $ \mathcal{T}_t' $, and $ \mathrm{rank}(t) $ is the position of test case $ t $ in the ranked test suite $ \mathcal{T}_t' $.
$ p $ is the number of faults detected by $ \mathcal{T}_t' $ divided by the number of faults contained in the full test suite $ \mathcal{T}_t $.
The NAPFD value ranges from 0 to 1, with a higher value indicating better prioritization or selection~\cite{Li2021}.

To evaluate the feedback time for the developer in case of a faulty code change, we consider two additional metrics.
First, we use the number of test cases executed before the first failure normalized to the size of $ \mathcal{T}' $, denoted as the normalized failure rank (NFR), to measure the ranking performance.
Further, we introduce the normalized time to fail (NTTF), which measures the execution time until the first failing test case normalized to the full resource time budget.
Smaller values indicate better performance for both metrics.

To evaluate the effectiveness of our post-submit test selection, we assessed how many developer-relevant transitions were identified by our approach.
Furthermore, we investigated how many CI cycles it took to select all information-gaining transitions at least once.

\subsection{Threats to Validity}

\subsubsection{External Validity}
The generalizability of our findings is limited by the fact that our study was conducted within the specific industrial context of BMW.
While our codebase and CI infrastructure are representative of large-scale software systems in the automotive domain, the results may not be directly applicable to other domains or development environments with different characteristics. To enhance external validity, we have provided detailed descriptions of our industrial context, codebase, and CI infrastructure, enabling researchers and practitioners to assess the applicability of our approach to their specific contexts.

Furthermore, our study focused on a monorepo codebase written primarily in C++ and Python, using specific tools and frameworks (e.g., Zuul, Bazel). The applicability of our approach to other programming languages, build systems, or CI tools may require further investigation and adaptation.

\subsubsection{Construct Validity}
We mitigated threats to construct validity by carefully selecting and justifying the evaluation metrics used in our study. While the NAPFD metric is widely used for evaluating test case prioritization techniques, it may not fully capture all aspects of feedback latency and resource consumption in a CI pipeline. To address this limitation, we complemented the NAPFD metric with additional metrics, such as the NFR and the NTTF, to provide a more comprehensive evaluation of feedback latency.

For post-submit test selection, we used metrics like the number of identified transitions and the number of CI cycles required to select all information-gaining transitions. While these metrics may not comprehensively reflect the effectiveness and efficiency of our approach in all scenarios, they provide a reasonable approximation based on our industrial context.

Our study assumes that each test case reveals only one fault of the same severity. While this assumption may not hold true in all cases, as test cases can potentially detect multiple faults or faults of varying severities, it is a common simplification used in test case prioritization studies.

\subsubsection{Internal Validity}
We mitigated threats to internal validity by implementing mechanisms to detect and filter out corrupted or incomplete data in our data collection and processing pipeline. However, there may still be undetected issues that could affect our results. Our approach also relies on the accuracy of the test execution logs, which may contain inaccuracies or inconsistencies that we could not fully control.

Changes in the codebase, test infrastructure, or development practices during the data collection period may have impacted the test execution behavior, leading to biased or inconsistent results. To mitigate this threat, we carefully monitored any significant changes that occurred during the study period.


\section{Results}


This section outlines the performance reached by our \deepq-based regression test optimization, which we denote as PR-DQL, against the baselines.
For PR-DQL, COLEMAN, and ROCKET we report their average scores over all cycles within the datasets.
For the RANDOM baseline, we show the averaged scores over \num{10000} repetitions.
First, we evaluated their effectiveness when ranking and selecting tests running in pre-submit pipelines based on their failure-likelihood (RQ1).
The second part consists of the results for selecting developer-relevant transitions in post-submit pipelines (RQ2).


\subsection{RQ1: Pre-Submit Test Target Selection}

\begin{table}
    \caption{Experimental results on a $50\,\%$ resource budget.}
  \begin{center}
      \begin{tabular}{ lrrr }
        \toprule
        & \textbf{NAPFD} $\boldsymbol{\uparrow}$ & \textbf{NFR} $\boldsymbol{\downarrow}$ & \textbf{NTTF} $\boldsymbol{\downarrow}$ \\
        \midrule
         \textbf{\costrank} & $ \mathbf{0.71 \pm 0.31} $ & $ \mathbf{0.16 \pm 0.22} $ & $ \mathbf{0.16 \pm 0.24 }$\\
          \rnfail & $ 0.70 \pm 0.32 $ &  $ 0.17 \pm 0.22 $ & $ 0.19 \pm 0.25 $ \\
      \bottomrule
      \end{tabular}
      \label{table:checkFeatureEvaluation}
  \end{center}
\end{table}

\begin{table}
  \caption{NAPFD $\boldsymbol{\uparrow}$ and NFR $\boldsymbol{\downarrow} $ for prioritization ($\boldsymbol{100\,\%}$ budget).}
\begin{center}
    \begin{tabular}{ lrrr }
     \toprule
      & \textbf{NAPFD $\boldsymbol{\uparrow}$} & \textbf{NFR $\boldsymbol{\downarrow} $} & \textbf{NTTF $\boldsymbol{\downarrow} $}\\
     \midrule
     \textbf{PR-DQL} & $ \mathbf{0.75 \pm 0.25} $ & $ \mathbf{0.17 \pm 0.22} $ & $ \mathbf{0.19 \pm 0.25}$ \\
     COLEMAN & $0.72 \pm 0.27$ & $ 0.19 \pm 0.24 $ & $0.20 \pm 0.18$\\
     ROCKET & $ 0.58 \pm 0.34 $ & $ 0.30 \pm 0.36 $ & $0.34 \pm 0.38$ \\
     RANDOM & $ 0.48 \pm 0.21 $ & $0.31 \pm 0.27$ & $0.31 \pm 0.27$ \\
     \bottomrule
    \end{tabular}
    \label{table:checkComparison}
\end{center}
\end{table}

\begin{table}[t]
  \begin{center}
    \caption{NAPFD $\boldsymbol{\uparrow}$ for prioritization and selection.}
      \begin{tabular}{ rll }
       \toprule
        & \textbf{NAPFD $\boldsymbol{\uparrow}$}\\
       \midrule
       \textbf{Budget: $10\,\%$} & \\
       \hline
       \textbf{PR-DQL} &  $ \mathbf{0.53 \pm 0.42 }$\\
       COLEMAN & $ 0.52 \pm 0.41 $ \\
       ROCKET &  $ 0.23 \pm 0.37 $ \\
       RANDOM & $ 0.13 \pm 0.25 $ \\
       \hline
       \textbf{Budget: $50\,\%$} \\
       \hline
       \textbf{PR-DQL}&  $ \mathbf{0.71 \pm 0.31}$ \\
       COLEMAN & $0.68 \pm 0.31$ \\
       ROCKET & $0.44 \pm 0.41$ \\
       RANDOM & $ 0.39 \pm 0.29 $ \\
       \hline
       \textbf{Budget: $80\,\%$} \\
       \hline
       \textbf{PR-DQL}&  $\mathbf{0.71 \pm 0.28} $\\
       \textbf{COLEMAN} & $ \mathbf{0.71 \pm 0.27}$ \\
       ROCKET & $ 0.52 \pm 0.39 $ \\
       RANDOM & $ 0.47 \pm 0.24 $ \\
       \bottomrule
      \end{tabular}
      \label{table:checkBudgetComparison}
  \end{center}
\end{table}

\Cref{table:checkFeatureEvaluation} compares the performance of our novel \costrank reward function against its non-cost-aware baseline (\rnfail) using a time budget of \perc{50} (we saw similar results for different resource budgets). \costrank, which takes into account both the rank and cost of each test case, improves the performance compared to the \rnfail reward function in all categories.
The improvement of the average NFR and NTTF values show that incorporating the rank and execution time in the reward function improves feedback latencies.
The high complexity of predicting test priorities is reflected in the high standard deviation of the performance metrics.

Based on the above results we compare our \deepq-based (referred to as PR-DQL) approach using the \costrank reward-function against the mentioned baselines.
Regarding COLEMAN, we chose the best proposed configuration for comparison~\cite{Lima2022}.
We started comparing the ranking performances and set the time budget to \perc{100}, as failing tests may not be contained in the selected test suite otherwise. \Cref{table:checkComparison} shows the mean NAPFD, NFR, and NTTF of the different approaches on our pre-submit dataset.
PR-DQL outperforms all other approaches on all three metrics.
It achieves the highest NAPFD of $0.75 \pm 0.25$, indicating that on average, it detects failures earlier. It also has the lowest NFR of $0.17 \pm 0.22$, meaning that the first failing test case occurs at $17\%$ of the test suite execution on average. Additionally, PR-DQL has the lowest NTTF of $0.19 \pm 0.25$, implying that it finds the first failing test case within the first $19\%$ of the test suite execution on average.
In comparison, the COLEMAN approach achieves an NAPFD of $0.72 \pm 0.27$ and an NFR of $0.19 \pm 0.24$, slightly lower than PR-DQL.
ROCKET and RANDOM perform substantially worse than PR-DQL yielding notably lower NAPFD values
and higher NFR values%
.

The lower standard deviations of PR-DQL on NAPFD are important, as stability makes them more  suitable for industry integration. Overall, these results demonstrate that our proposed \deepq-based approach with the \costrank reward function improves the ranking and prioritization of test cases compared to the baselines, leading to earlier fault detection and more efficient test execution.

As we also want to reduce the drag on our CI to further minimize feedback latencies, we compared our approach against the baselines using different time budgets.
We used several budgets, ranging from $10\,\%$ to $80\,\%$ of the total test suite execution time, to investigate the trade-off between execution time and fault detection capabilities. \cref{table:checkBudgetComparison} shows the NAPFD values for these different budgets.

For all budget levels, PR-DQL outperforms or matches the baselines, with the difference being more pronounced at lower budgets. At a $10\,\%$ budget, PR-DQL achieves an NAPFD of $0.53 \pm 0.42$, substantially higher than COLEMAN ($0.52 \pm 0.41$), ROCKET ($0.23 \pm 0.37$), and RANDOM ($0.13 \pm 0.25$). This advantage persists at higher budgets, with PR-DQL consistently achieving the highest NAPFD values. At an $80\,\%$ budget, PR-DQL and COLEMAN perform equally well with an NAPFD of $0.71$, but both outperform ROCKET ($0.52 \pm 0.39$) and RANDOM ($0.47 \pm 0.24$).

These results demonstrate the effectiveness of PR-DQL in prioritizing and selecting the most relevant test cases, even under tight time constraints. By leveraging the \costrank reward function and the \deepq learning approach, PR-DQL can efficiently identify the most fault-revealing test cases, leading to earlier fault detection and reduced test execution time. This capability is particularly valuable in continuous integration environments, where minimizing feedback latencies is crucial for efficient software development workflows.

\summary{1}{
  We evaluated our proposed \deepq-based approach (PR-DQL) using the \costrank reward function against several baselines on our pre-submit dataset. PR-DQL outperformed all baselines across multiple metrics, including NAPFD ($\boldsymbol{\uparrow}$), NFR ($\boldsymbol{\downarrow}$), and NTTF ($\boldsymbol{\downarrow}$), demonstrating its effectiveness in prioritizing and selecting the most fault-revealing test cases. PR-DQL maintained its advantage over baselines even under tight time constraints, achieving the highest NAPFD values across different budget levels.
 The lower standard deviations of PR-DQL on NAPFD underlined its stability and suitability for industry integration.
  These results demonstrate the effectiveness of PR-DQL with the \costrank reward function in improving test case prioritization and selection, leading to earlier fault detection and reduced feedback latencies in CI environments.
}

\subsection{RQ2: Post-Submit Test Target Selection}

This section evaluates the capability of the proposed reward functions (\ie \rnchange and \costcrank) to choose pertinent test targets for post-submit pipelines by comparing different reward functions using different time budgets against the random baseline. In this context, we consider all test cases with a change in the execution result and a positive information gain for developers as relevant.
We do not regard transitions due to flakiness as relevant.

\begin{figure}
    \centering
    \scalebox{1}{\input{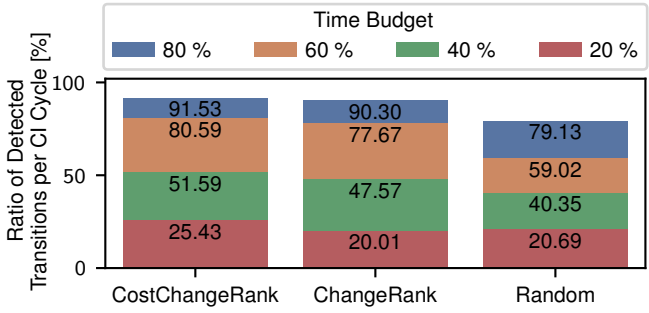}}
    \caption{Percentage of test cases executed in each CI cycle that reveal a relevant test result transition.}
    \label{fig:ExecutionResults}
\end{figure}

\Cref{fig:ExecutionResults} shows the percentage of test cases executed in each CI cycle that reveal a relevant test result transition, providing useful information gain for developers. The results are presented for three different approaches: \costcrank, \rnchange, and the random baseline. Four different time budgets are considered: \perc{80}, \perc{60}, \perc{40}, and \perc{20}.
With a \perc{80} time budget, the \costcrank reward function selected \perc{91.53} of the test cases that revealed relevant transitions, outperforming the \rnchange reward function (\perc{90.3}) and the random baseline (\perc{79.13}). As the time budget decreased, the performance gap between \costcrank and the other approaches widened.
At a \perc{60} time budget, \costcrank selected \perc{80.59} of the relevant test cases, while \rnchange selected \perc{77.67}, and the random baseline selected only \perc{59.02}. With a \perc{40} time budget, \costcrank identified \perc{51.59} of the relevant transitions, compared to \perc{47.57} for \rnchange and \perc{40.35} for the random approach.
Even under tight time constraints of a \perc{20} budget, \costcrank managed to select \perc{25.43} of the test cases with relevant transitions, outperforming \rnchange (\perc{20.01}) and the random baseline (\perc{20.69}). These results demonstrate the effectiveness of the \costcrank reward function in selecting pertinent test cases that provide valuable information gain for developers, particularly in scenarios with limited computational resources.

\begin{figure}
    \centering
    \scalebox{1}{\input{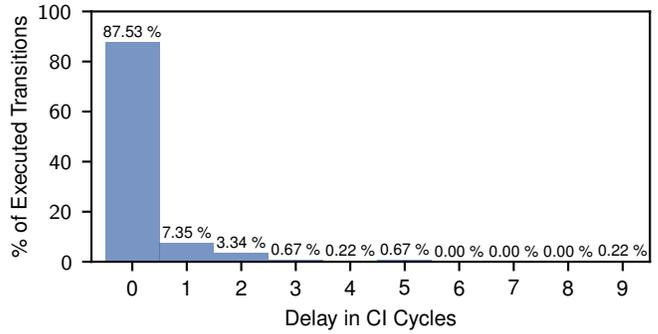}}
    \caption{Number of test selection CI cycles required until a relevant test result transition is detected for a resource budget  of $\boldsymbol{50\,\%}$.}
    \label{fig:ExecutionDelay}
\end{figure}

\Cref{fig:ExecutionDelay} illustrates the number of CI cycles required until a relevant test result transition is detected when using the \costcrank approach with a resource budget of \perc{50}.
Since code changes merge into the main branch before the post-pipeline execution, test targets that were initially missed can reveal transitions in later executions of subsequent code changes.
The figure shows that over \perc{87} of transitions in the dataset are detected immediately without any additional delay. Moreover, over \perc{98} of transitions are detected within three subsequent CI cycles. Remarkably, all transitions are detected within ten CI cycles. As these test cases execute in post-pipeline jobs, immediate feedback for developers is not as critical. A delay of up to ten post-submit cycles thus seems negligible, offering a suitable trade-off between CI resource costs and time delay.

The \costcrank reward function incorporates execution time, affecting the estimated priority value. Generally, test targets with shorter durations execute more frequently than longer ones. However, the evaluation shows that in some exceptions, longer-running test cases in a specific duration region execute more frequently as well. This indicates that the algorithm strikes a balance between scheduling multiple smaller test cases versus fewer but longer ones.

\summary{2}{
  We compared our novel reward function \costcrank both against a RANDOM baseline and a non-cost-aware reward function (\rnchange) under different time budgets.
  The \costcrank reward function outperforms \mbox{\rnchange} and the RANDOM baseline in selecting test targets that reveal relevant transitions (\ie non-flaky regressions and bug fixes), providing valuable information  for developers while reducing load on our CI. The performance gap between \costcrank and the compared baselines widens as the time budget decreases.
  With a \perc{50} time budget, over \perc{87} of transitions are detected immediately by our approach, and over \perc{98} within three CI cycles. All transitions are detected within ten cycles, offering a suitable trade-off between CI resource costs and test feedback delay.
}


\section{Discussion and Application}

Developing software in large-scale monorepos efficiently requires effective regression test optimization techniques.
Current research efforts often focus on predicting test case failure likelihood, employing complex features such as coverage information or code complexity metrics.
However, these approaches can introduce significant overhead and dependencies, hindering their scalability and adoption in industry-scale environments.
Our approach takes a different path, recognizing the importance of a language-agnostic, lightweight, and pipeline-aware regression testing optimization solutions.
We acknowledge that different stages of our CI system have distinct characteristics and objectives.
Our proposed reward functions consider pipeline characteristics and test costs, ensuring that the prioritization and selection of test targets are informed by the unique objectives of each pipeline.
This pipeline-aware design puts the developer experience first, asking which test execution is most relevant at different CI stages and prioritizing tests accordingly, while aiming to minimize resource consumption.
Our results underline the importance of tailoring quality assurance optimizations to the specific CI pipeline they are intended to operate in, as
a one-size-fits-all approach may not be effective in addressing unique challenges and requirements of different software integration stages.
Applying test selection in CI can lead to incorrect omissions, i.e., missed test failures.
In pre-submit pipelines, this can cause test failures in later CI stages (such as our gate pipeline) or block the integration of future changes.
In post-submit pipelines, they lead to incorrect culprit assignments.
As both gate failures and  regression misattributions are serious issues, safeguard mechanisms need to be introduced before deploying test selection to production.
We plan to monitor gate failures caused by falsely omitted tests and introduce an automated bisecting mechanism to ensure correct culprit finding.
%

Safety and certification requirements might pose a challenge to the adoption of test selection, but as we do not apply test selection to release testing this is not an issue in our scenario: We always execute all tests before releasing our software.

\section{Related Work}

Researchers have studied different approaches for test case prioritization and selection. ROCKET is a heuristic-based prioritization approach that orders test cases based on historical test results and execution times~\cite{Marijan2013}. It computes the priority using a failure matrix containing historical execution results, weighted by the execution date. Microsoft proposes a test prioritization and selection solution based on a lightweight statistical model, using file changes in a PR and historical test execution results as input features~\cite{Mehta2021}. Given our constantly changing CI, such heuristics might need to be adopted over their lifespan, requiring additional effort for continuous adaptation.

TCP-Net is a deep neural network (DNN) for test case prioritization that combines file and test case-related features using deep fusion. It encodes textual features using a BoW model together with PCA~\cite{Abdelkarim2022}. DeepOrder utilizes a DNN for ranking test cases based on the historical record of test executions. DeepOrder employs a DNN with a few hidden layers and is developed for system-level test cases, making it independent of the source code~\cite{Sharif2021}. While DeepOrder ranks system-level tests, we require a system to rank all sort of tests ranging from unit- to simulation-tests spanning across our CI.

In the domain of test case selection, where only a subset of available tests is executed, the inputs (historical execution results) depend on the decisions made by the model in prior cycles, requiring incremental learning~\cite{Spieker2017}. Additionally, the CI environment may change due to renamed or deleted tests, which would necessitate retraining~\cite{Abdelkarim2022, Spieker2017, Bertolino2020}. To address these drawbacks, recent research in test case selection and prioritization~\cite{Spieker2017, Bertolino2020, Lima2022, Bagherzadeh2022} has adapted \rl, which is well-suited for dynamic environments as it progressively adapts and improves~\cite{Spieker2017}. Therefore, we chose to use a \rl approach as well.

State-of-the-art \rl-based algorithms include RETECS~\cite{Spieker2017} and COLEMAN~\cite{Lima2022}. RETECS selects and prioritizes test cases according to their duration, time of last execution, and failure history using \rl together with artificial neural networks. COLEMAN is a test case prioritization approach for CI environments based on a multi-armed bandit with combinatorial and volatile characteristics~\cite{Lima2022}. Compared to RETECS, it only requires minimal information (historical failures) to prioritize test cases and does not need duration or last execution time~\cite{Lima2022, Spieker2017}.

Researchers have identified the necessity to differentiate regression testing depending on the CI pipeline~\cite{Elbaum2014}. They propose a heuristic-based test case selection algorithm for the pre-submit phase of testing and a prioritization algorithm for the post-submit phase. Further studies have examined differentiating between a PR validation stage and a comprehensive validation stage~\cite{Mehta2021}. However, current research lacks this differentiation and generally focuses on pre-submit characteristics. Therefore, this work proposes a \rl-based solution and investigates different pipeline characteristics, adapting the algorithm based on them.


\section{Conclusions}

Scaling Continuous Integration to massive, monolithic repositories requires regression test optimization techniques that are apt for such fast changing, multi-language, multi-pipeline environments.
Since many existing techniques are not able to operate in such a context, we developed a lightweight regression test selection and prioritization approach that is based on \deepq, a \rl technique, with pipeline-specific reward functions:
For pre-submit pipelines, it creates test prioritizations, optimizing against fast developer feedback by considering the average test durations.
For post-submit pipelines, it creates test selections, aiming to minimize resource consumption while still providing developers with relevant feedback by taking potential test flakiness into account and predicting test transitions instead of test failures.

An evaluation on a large-scale industry dataset showed that our approach is able to outperform existing techniques.
In the pre-submit pipeline, it was to able to schedule the first failing test within the first 16\% of test cases on average.
In the post-submit pipeline, it was able to select 87\% of developer-relevant test transition using a 50\% time budget, and detect 99.78\% of relevant test transition within five CI cycles.

We hope that our work inspires more research on regression test optimization in large-scale CI environments.



\bibliographystyle{IEEEtran}
\balance
\bibliography{main.bbl}

All online resources accessed on 2025--01--02.

\end{document}